\documentclass [11pt, a4paper, notitlepage] {article}

\usepackage [polish, english] {babel}
\usepackage [cp1250] {inputenc}
\usepackage [T1] {fontenc}
\usepackage {polski}
\usepackage {graphicx, epsfig}
\usepackage {amsmath} 

\begin{document}

\bibliographystyle{plane}

\title{Inflation and Preheating in Supergravity with MSSM Flat Directions}
\date{}
\author{Anna Kami\'{n}ska and Pawe{\l} Pacho{\l}ek \\
Institute of Theoretical Physics, Faculty of Physics, \\
University of Warsaw, Ho{\.z}a 69, Warsaw, Poland}

\maketitle

\begin{abstract}
Motivated by a recent discussion about the role of flat directions, a typical feature of supersymmetric models, in the process of particle production in the early universe a consistent model of inflation and preheating in supergravity with MSSM fields has been built. It is based on a model proposed by M. Kawasaki, M. Yamaguchi and T. Yanagida. In the inflationary stage, the flat directions acquire large vacuum expectation values (VEVs) without spoiling the background of slow-roll, high-scale inflation consistent with the latest WMAP5 observational data. In the stage of particle production, naturally following inflation, the role of flat direction large VEVs depends strongly on effects connected with the supergravity framework and non-renormalizable terms in the superpotential, which have been neglected so far in the literature. Such effects turn out to be very important, changing the previous picture of preheating in the presence of large flat direction VEVs by allowing for efficient preheating from the inflaton. \\
\end{abstract}

\section{Introduction}

Inflation was introduced as a natural and simple way of solving the problems of classical cosmology - the initial conditions problem (or the flatness and horizon problems) and the explanation of the origin of primordial density fluctuations \cite{guth, linde1, albrecht, guth1, mukhanov1, mukhanov2, colb, mukhanov}. The easiest way to obtain inflation is by introducing a single scalar inflaton field with a slowly evolving vacuum expectation value \cite{linde1, albrecht}. In order to obtain a proper period of Big Bang Nucleosynthesis however, one has to end inflation by particle production. The process of reheating must connect the inflaton sector to the observable sector \cite{albrecht1, abbott, dolgov}.\newline
In order to properly describe inflation and particle production one has to consider the underlying theory of particles and interactions. Supersymmetry is one of the most promising extensions of the Standard Model (SM) \cite{fayet, nilles, haber, bailin, ramond, binetruy}, and it has triggered a search for supersymmetric models of inflation and reheating. One of the typical features of supersymmetric extensions of the SM is the presence of flat directions \cite{gherghetta} - directions in field space, along which the scalar potential identically vanishes in the limit of unbroken global supersymmetry. Due to large quantum fluctuations or the classical evolution of fields during inflation flat directions can easily acquire large VEVs \cite{dine, kasuyaa}. Therefore, there is a natural question about the role of such large VEVs in the process of particle production.\newline
It was postulated in ref. \cite{mazumdar}, that large flat direction VEVs influence the process of particle production by blocking preheating from the inflaton - the phase of rapid, non-perturbative inflaton decay. In ref. \cite{mazumdar} a simple toy model was proposed
\begin{equation}
V\supset{}A\varphi^{2}\chi^{2}+B{}m\varphi\chi^{2}+C\alpha^{2}\chi^{2},
\label{spi}
\end{equation}
where $\varphi$ is the inflaton field, $\alpha$ parameterizes the flat direction and $\chi$ represents the inflaton decay products (in this model a direction in Higgs fields has been considered). Then, after mode decomposition of the field $\chi$, the energy of the mode with momentum $k$ is given by:
\begin{equation}
\omega_{k}^{2}=k^{2}+2A\left\langle\varphi\right\rangle^{2}+2B{}m\left\langle\varphi\right\rangle+2C\left\langle\alpha\right\rangle^{2}.
\end{equation}
In general, non-adiabatic production of particles $\chi_{k}$ is efficient only when $\omega_{k}$ changes non-adiabatically
\begin{equation}
|\tau|\equiv\left|\frac{\dot{\omega}}{\omega^{2}}\right|>1\leftrightarrow{}preheating,
\label{adiabat}
\end{equation}
where the adiabaticity parameter $\tau$ is introduced. During classical preheating $\omega_{k}$ is dominated by the inflaton VEV and changes non-adiabatically due to inflaton oscillations. In the presence of flat directions however, $\omega_{k}$ could be dominated by the large VEV of the flat direction. If this VEV changes very slowly in comparison with the evolution of the inflaton VEV, non-perturbative production of $\chi$ particles is effectively blocked.\newline
However, as was pointed out in ref. \cite{olive}, blocking of preheating from the inflaton does not occur when non-perturbative production of particles from the flat direction itself is possible. Then the initially large VEV of the flat direction decreases rapidly, unblocking preheating from the inflaton. In ref. \cite{olive} a method was introduced of calculating the amount of particles produced non-perturbatively from the flat direction, due to non-adiabatic changes of the mass matrix eigenvectors and eigenvalues related to quantum fluctuations around the flat direction. This led to a discussion (see refs \cite{allahverdi, basboll1, basboll, olive1}) about whether non-perturbative decay of flat directions and preheating from the inflaton is possible. The discussion was based on some general properties of flat directions in a global supersymmetry framework. It did not consider any specific model of inflation and did not propose any model of acquiring large VEVs by flat directions. Therefore it was difficult to study the whole issue and determine how the large VEVs of flat directions develop and evolve, and how they impact the process of inflation and particle production.\newline
The goal of our work is to construct a consistent model of inflation and particle production in a realistic supersymmetric extension of the Standard Model, and consider in this specific model the behavior of MSSM flat directions. Therefore a realistic chaotic inflation model with two representative flat directions is constructed. In order to be able to predict the evolution of flat direction VEVs it was decided to study the production of large flat direction VEVs by classical evolution during inflation, and therefore a potential for the flat direction is required. Following refs \cite{dine, kasuyaa} we adopt the supergravity framework with a non-minimal K\"{a}hler potential, which results in a potential for the flat direction with a time-evolving minimum at large VEVs during inflation. It also enables us to calculate the previously neglected influence of supergravity corrections for flat direction evolution during inflaton oscillations. We also consider the impact of existence of non-renormalizable terms, which has also been neglected so far. We find that these effects strongly influence the process of particle production by introducing efficient channels of non-perturbative particle production both from the flat direction and the inflaton. As a result the originally large flat direction VEVs are diminished, preheating from the inflaton is allowed and the energy density of the Universe is dominated by the inflaton decay products.

\section{Building the model}

The model considered in this paper, after neglecting all fields except for the inflaton, reduces to the simplest chaotic inflation model with the inflaton potential $V=m^{2}\varphi^{2}/2$. This property together with inflaton domination provides appropriate slow-roll inflation and a value of spectral index which is in agreement with the WMAP5 data \cite{komatsu}. Obtaining such a property in a model which is based on supergravity is not straightforward due to complicated supergravitational F-terms. A solution to this problem (the so called $\eta$-problem) was proposed by \cite{kawasaki} and is used in this paper. According to the solution, except for the chiral inflaton superfield $\Phi$ and the MSSM superfields, the model contains one additional chiral superfield $X$.\newline Further consideration is restricted to scalar fields and the same symbol is used to denote both the chiral superfield and its complex scalar component. The following decomposition in real fields is used
\begin{equation}
\Phi=(\eta+i\varphi)/\sqrt{2},
\label{Phidec}
\end{equation}
\begin{equation}
X=xe^{i\beta}.
\label{Xdec}
\end{equation}
The field $\varphi$ plays the role of the inflaton.\newline
We follow \cite{kawasaki} in constructing the K\"{a}hler Potential $K$ and take
\begin{equation}
K\supset\frac{1}{2}(\Phi+\Phi^{*})^{2}+XX^{*}.
\label{Kh}
\end{equation}
% Wyjaniæ i opisaæ model Kawasaki (shift symmetry itp.)
The formula (\ref{Kh}) for the K\"{a}hler potential was obtained by \cite{kawasaki} as follows. The first step was to introduce a Nambu-Goldstone-like shift symmetry of the inflaton $\alpha$
\begin{equation}
\Phi \rightarrow \Phi +iCM,
\label{shift}
\end{equation}
where $C$ is a dimensionless real parameter. A K\"{a}hler potential which is invariant under this symmetry and the additional $U(1)_{R}\times Z_{2}$ symmetry must have the general form \cite{kawasaki}
\begin{equation}
K(\Phi,\Phi^{*},X, X^{*}) = K[(\Phi+\Phi^{*})^{2},XX^{*}].
\label{Kgen}
\end{equation}
The formula (\ref{Kh}) is just the lowest order term in the general expansion of the formula (\ref{Kgen}). 
It can be easily seen that a theory with an exact Nambu-Goldstone-like shift symmetry has no potential for the inflaton $\varphi$. Therefore this symmetry has to be broken, but not in the K\"{a}hler potential in order to avoid the $\eta$-problem. Following \cite{kawasaki}, we introduce a shift symmetry breaking term in the superpotential $W$
\begin{equation}
W\supset mX\Phi.
\label{Wh}
\end{equation}
It gives mass $m$ to the inflaton $\varphi$.\newline
There are two MSSM-flat directions considered in this paper: a direction $H_{u}H_{d}$ in Higgs fields (only D-flat) and a direction $u_{i}d_{j}d_{k}$ in squark fields, where indexes $i$, $j$ and $k$ are some family indexes ($k\neq j$). Let $\chi$ be the complex scalar field that parametrize the $H_{u}H_{d}$ direction
\begin{equation}
H_{d}=\frac{1}{\sqrt{2}}\left(\begin{array}{cc} 
      \chi \\ 0 \end{array}\right),\ H_{u}=\frac{1}{\sqrt{2}}\left(\begin{array}{cc} 
       0 \\\chi \end{array}\right).
\label{chiflat}
\end{equation}
Let $\alpha$ be the complex scalar field which parametrizes the $u_{i}d_{j}d_{k}$ direction
\begin{equation}
u^{\beta}_{i}=d^{\gamma}_{j}=d^{\delta}_{k}=\frac{1}{\sqrt{3}}\alpha.
\label{phiflat}
\end{equation}
In equation (\ref{phiflat}) $\beta \neq \gamma \neq \delta \neq \beta$ are fixed color indexes. The components of fields $u_{i}$, $d_{j}$ and $d_{k}$ with other color indexes are equal to zero.
It is convenient to decompose the complex fields $\chi$ and $\alpha$ into real fields in the following way
\begin{equation}
\chi=ce^{i\kappa},
\label{chidec}
\end{equation}
\begin{equation}
\alpha=\rho e^{i\sigma}.
\label{phidec}
\end{equation}\newline
The full K\"{a}hler potential is
\begin{equation}
K=\frac{1}{2}(\Phi+\Phi^{*})^{2}+XX^{*}+K_{MSSM}+K_{NM},
\label{K}
\end{equation}
where $K_{MSSM}$ is a standard minimal K\"{a}hler potential and $K_{NM}$ is a non-minimal part of the form
\begin{equation}
\begin{array}{ll} K_{NM}=\frac{a}{M^{2}_{Pl}}XX^{*}\cdot \\ \cdot(H_{u}^{+}H_{u}+H_{d}^{+}H_{d}+u_{i}^{+}u_{i}+d_{j}^{+}d_{j}+d_{k}^{+}d_{k}).\end{array}
\label{K_NM}
\end{equation}
Here $M_{Pl}$ is the Planck mass and $a$ is a dimensionless parameter. The existence of couplings like $K_{NM}$ is guaranteed in the presence of Yukawa couplings, since they are necessary counterterms for operators generated by loop diagrams \cite{dine, bagger, gaillard}. Terms in $K_{NM}$ cause the existence of minima in the scalar potential for both flat directions, which are of the order of $M_{Pl}$. Therefore flat directions can naturally aquire large VEVs of the order of $M_{Pl}$ by falling into these minima.
\newline
For the superpotential we take
\begin{equation}
W=mX\Phi+2hXH_{u}\cdot{}H_{d}+W_{MSSM}+W_{NR}.
\label{W}
\end{equation}
The term $2hXH_{u}\cdot{}H_{d}$ is one of only two possible renormalizable couplings between the field $X$ and MSUGRA fields. The second one is $2h'XH_{u}\cdot{}L$. These two couplings cannot coexist in the model unless R-parity is broken. We assume that the inflaton does not couple to MSSM fields in the K\"{a}hler potential and in the superpotential to avoid strong deviations from the slow-roll inflation regime with inflaton domination. 
The term $W_{MSSM}$ is the standard MSSM superpotential, given by
\begin{equation}
W_{MSSM}=\mu{}H_{u}\cdot{}H_{d}+\lambda_{u}^{lm}H_{u}\cdot{}Q_{l}u_{m}+\lambda_{d}^{lm}H_{d}\cdot{}Q_{l}d_{m}.
\label{W_MSSM}
\end{equation}
The last term $W_{NR}$ is a non-renormalizable part of the superpotential and it has the following form
\begin{equation}
W_{NR}=\frac{\lambda_{\chi}}{M_{Pl}}\left(H_{u}\cdot{}H_{d}\right)^{2}+\frac{3\sqrt{3}\lambda_{\alpha}}{M_{Pl}}\left(u_{i}d_{j}d_{k}\nu_{R}\right).
\label{W_NR}
\end{equation}
Here $\lambda_{\chi}$ and $\lambda_{\alpha}$ are dimensionless constants and $\nu_{R}$ is a right-handed neutrino of any given generation. Two non-renormalizable terms contained in $W_{NR}$ are the only terms of 4th order in the fields, which may be relevant for the evolution of VEVs of the two chosen flat directions. The term $3\sqrt{3}\lambda_{\alpha}'\left(u_{i}u_{j}d_{k}\nu_{R}\right)/M_{Pl}$ 
gives no contribution to the scalar potential, unless one additionally considers VEVs of some other flat directions.\newline
Terms in $W_{NR}$ modify minima for flat directions, which are shifted away from zero in the scalar potential due to terms in $K_{NM}$. The minima are no longer close to $M_{Pl}$ when coupling constants $\lambda_{\chi}$ and $\lambda_{\alpha}$ are sufficiently large. Moreover, they evolve in time during inflation until they reach their final position at zero at the end of the inflaton oscillations.\newline
The soft SUSY-breaking terms in the scalar potential have negligible effects on our results.\newline
Our initial conditions are set for the time which corresponds to about 100 e-folds before the end of inflation, since only this period is essential for preheating. Initial values for the real fields $\varphi$, $\eta$, $x$, $\rho$ and $c$ do not require fine tuning. For the inflaton field $\varphi$ the only condition which needs to be satisfied is $\varphi_{0} > M_{Pl}$. It ensures that the number of e-folds is greater than 75, which is needed for inflation to solve the horizon problem and the flatness problem \cite{mukhanov}. We took the initial value $\varphi_{0}= 4M_{Pl}$ to consider only the last 100 e-folds before the end of inflation. Fields $\eta$, $x$, $\rho$ and $c$ should be initially smaller than $M_{Pl}$ because they are present in K\"{a}hler potential, especially in the exponential factor $\exp(\frac{K}{M_{Pl}^{2}})$ in the F-terms. Therefore initially large values of any of those fields should decrease rapidly and the VEVs of the fields $\eta$, $x$, $\rho$ and $c$ should stay confined naturally below the Planck scale during inflation. In particular the VEV of field $\eta$ falls to zero very quickly and, as we have checked numerically, does not have any noticable influence on the evolution of other fields. Therefore in our final calculations we simply put $\eta_{0}=0$. The evolution of the field $x$ has also almost no influence on the evolution of other fields, so we set for it a quite arbitrary initial value $x_{0}=0.01M_{Pl}$. Fields $\rho$ and $c$ which are absolute values for complex flat direction fields $\alpha$ and $\chi$ are initially taken to be of the order of the Hubble parameter $H$, which is also the order of initial quantum fluctuations for those fields during inflation. In the slow-roll regime with inflaton domination, we have $H_{0} \approx \sqrt{8\pi m^{2}\varphi_{0}^{2}/6M_{Pl}^{2}}$. After setting $m=10^{-6}M_{Pl}$ as in the simplest model of slow-roll inflation consistent with WMAP data we get $H_{0}\approx 8 \cdot 10^{-6}M_{Pl}$. Therefore we take $\rho_{0} =c_{0}=10^{-5}M_{Pl}$.
 
\section{Classical evolution of fields}

In order to study the process of acquiring large VEVs by flat directions one has to consider the classical evolution of fields during inflation. A classical description is possible due to the slow-roll character of the evolution. At the end of inflation, excitations around VEVs of the fields will be considered in order to determine the impact of large flat direction VEVs on the process of particle production.\newline
The classical evolution is determined by the equations of motion derived from the supergravity Lagrangian once the initial conditions have been set. The inflaton equation of motion has the simple form
\begin{equation}
\ddot{\varphi}+3H\dot{\varphi}+V,_{\varphi}=0.
\end{equation}
During inflaton domination the main contribution to the scalar potential V is of the form $1/2\;m^{2}\varphi^{2}$, which provides a standard chaotic inflation background. Due to the shift symmetry the K\"{a}hler potential does not depend on $\varphi$, and so the supergravity coefficient $e^{K}$ does not contain the inflaton field. This solves the $\eta$-problem and allows for high-scale inflation. Since only the last 80-100 e-folds of inflation have any observable consequences for the evolution of the Universe, the initial value of the inflaton VEV was chosen in a way that allows the study of this period of inflation. Due to the absence of any coupling of the inflaton with the $H_{u}H_{d}$ and $udd$ directions in the K\"{a}hler potential (because of the shift symmetry) the evolution of the inflaton is largely independent of the evolution of the flat directions. Because of this property the process of acquiring large VEVs by $udd$ or $H_{u}H_{d}$ directions will not spoil inflation.\newline
The equations of motion for X, $udd$ and Higgs fields are more complicated due to the non-minimal form of the K\"{a}hler potential for these fields. One can expect however that the field VEVs will evolve toward a minimum of the scalar potential. The scalar potential in supergravity is complicated as well. However one can observe that all the field VEVs except for the inflaton VEV are naturaly confined below the Planck scale due to the $e^{K}$ factor in the scalar potential. This factor becomes dominant in the scalar potential at the Planck scale. Therefore the scalar potential for all fields except the inflaton rises steeply at the Planck scale as $exp(field^{2})$. As a result one can expect $x,\ \rho,\ c$ to be less than unity and expand the scalar potential in these fields. For field $X=xe^{i\beta}$ the scalar potential has a minimum at zero. The term quadratic in field x is given by the following approximate expression (assuming inflaton domination and neglecting complex fields phases)
\begin{equation}
V\supset{}x^{2}\left(m^{2}+\frac{m^{2}\varphi^{2}}{2}\left(f(a)\rho^{2}+f(a)c^{2}+O\left[\rho^{2}c^{2},\;\rho^{3},\;c^{3}\right]\right)\right),
\end{equation}
where for simplicity we have set $M_{Pl}=1$ and $f(a)$ is positive for $a>0$. The evolution of $x$ is naturally confined to low VEVs due to the supergravity term $e^{K}$ which exponentially steepens the potential for field x at the Planck scale. The evolution of all other fields of the model is independent of the choice of initial conditions for $x$ and on the evolution of this field. The only role of field $X$ in the model is providing appropriate scalar potential for the inflaton and both flat directions.\newline
In the global MSSM without corrections coming from supergravity or non-renormalizable terms the scalar potential is independent of the flat direction $\alpha=\rho{}e^{i\sigma}$. Adding the corrections mentioned above creates a potential for the flat direction. A term quadratic in $\alpha$ in the scalar potential is created by supergravity effects and is sensitive to any non-minimal couplings in the K\"{a}hler potential. In the model presented in this paper the quadratic term mentioned above takes during inflaton domination (neglecting the complex fields phases) the following approximate form
\begin{equation}
V\supset\rho^{2}\;\frac{m^{2}\varphi^{2}}{2}\left(1-a+f(a)c^{2}+f(a)x^{2}+O\left[x^{2}c^{2},\;x^{3},\;c^{3}\right]\right).
\end{equation}
In the equation above the parameter $a$ describes the influence of the non-minimal coupling in the K\"{a}hler potential. In supergravity with a minimal K\"{a}hler potential the scalar potential has a global minimum at zero for the flat direction. In this case acquiring large flat direction VEVs due to classical evolution is impossible. The non-minimal coupling enables us to shift the minimum toward larger VEVs by an appropriate choice of $a>1$. Then the coefficient of the term quadratic in $\rho$ in the scalar potential becomes negative for $x,\;c<1$. For the purpose of numerical calculations we set $a=5$. The exact location of the minimum is determined by higher-order terms in $\rho$, which come from both supergravity and non-renormalizable terms. Supergravity alone stabilizes the minimum around $M_{Pl}$ due to the coefficient $e^{K}$. The presence of a non-renormalizable term scaled by $\lambda_{\alpha}$ shifts this minimum toward lower VEVs, changing the predicted flat direction VEVs at the end of inflation toward lower values. In order to study the effect of large flat direction VEVs on the process of particle production the value of $\lambda_{\alpha}$ should not exceed unity.\newline
Supergravity corrections and non-renormalizable terms have a similar effect on the potential for the $H_{u}H_{d}$ direction. The interplay between $\lambda_{\chi}$ and $\lambda_{\alpha}$ determines the differences between the evolution of the two flat directions ($H_{u}H_{d}$ and $udd$) under consideration. We study two specific scenarios
\begin{enumerate}
\item If $\lambda_{\alpha}\ll\lambda_{\chi}$ then the $udd$ direction VEV becomes large and the $H_{u}H_{d}$ direction VEV drops to zero during inflation, which corresponds to the scenario described in ref. \cite{mazumdar} in the limit of global supersymmetry without non-renormalizable terms. This case will enable us to study the predictions of ref. \cite{mazumdar} in a specific scenario and determine the impact of supergravity corrections and non-renormalizable terms, which has not been considered so far.
\item If $\lambda_{\alpha}\sim\lambda_{\chi}$ then both directions can acquire large VEVs during inflation and the impact of non-zero VEVs of the inflaton classical decay products on the process of particle production and blocking of preheating by flat direction large VEVs can be studied.
\end{enumerate}
By "large VEV" we mean a vacuum expectation value of the order $10^{-4}-1M_{Pl}$, which is large in comparison with the Hubble parameter $H\sim10^{-7}M_{Pl}$ at the beginning of inflaton oscillations. Since the acquired value of the $uud$ or $H_{u}H_{d}$ direction VEV is determined by the value of $\lambda_{\alpha}$ and $\lambda_{\chi}$ respectfully, in order to create large VEVs of those directions during inflation $\lambda_{\alpha}$ and $\lambda_{\chi}$ should not exceed unity. For the purpose of numerical calculations two sets of $\lambda$-parameters are considered
\begin{enumerate}
\item $\lambda_{\alpha}=10^{-7}$ and $\lambda_{\chi}=1$
\item $\lambda_{\alpha}=1$ and $\lambda_{\chi}=1$.
\end{enumerate}
In order to check if classical evolution can lead to large flat direction VEVs during inflation we have chosen small initial VEVs for both directions, which correspond to the average size of quantum fluctuations typical for the considered period of inflation ($\delta\alpha,\;\delta{}c\sim{}H$). In order to obtain numerical predictions in specific scenarios one has still to fix two free parameters. The choice of the inflaton mass $m\sim{}10^{-6}$ is natural because it implies a spectral index of energy density fluctuations consistent with WMAP observations. Following the arguments of ref. \cite{kawasaki1} we choose the inflaton coupling parameter $h$ to be small and of the order $h\sim{}10^{-5}$.\newline
The classical evolution of fields obtained numerically in the two scenarios mentioned above is presented below. It turns out that in both cases the evolution of field $\eta$ is irrelevant - the VEV of the $\eta$ field decreases rapidly to zero and does not influence the further evolution of the other fields. Henceforth we simplify our calculations by setting $\eta=0$. 

\subsection{$\lambda_{\alpha}=10^{-7}$ and $\lambda_{\chi}=1$}

\begin{figure}[h!]
\begin{minipage}[b]{0.47\linewidth}
\centering
\includegraphics[width=6 cm]{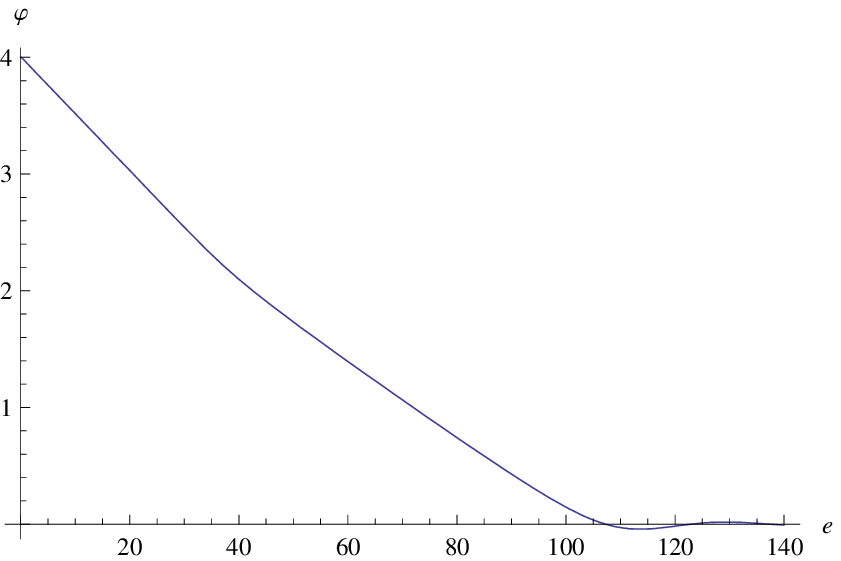}
\caption{Evolution of the inflaton field $\varphi$ during inflation. Values on vertical axes are expressed in Planck masses and time on horizontal axes is expressed in the approximate number of e-folds of inflation.}
\label{Pic1}
\end{minipage}
\hspace{0.5cm}
\begin{minipage}[b]{0.47\linewidth}
\centering
\includegraphics[width=6 cm]{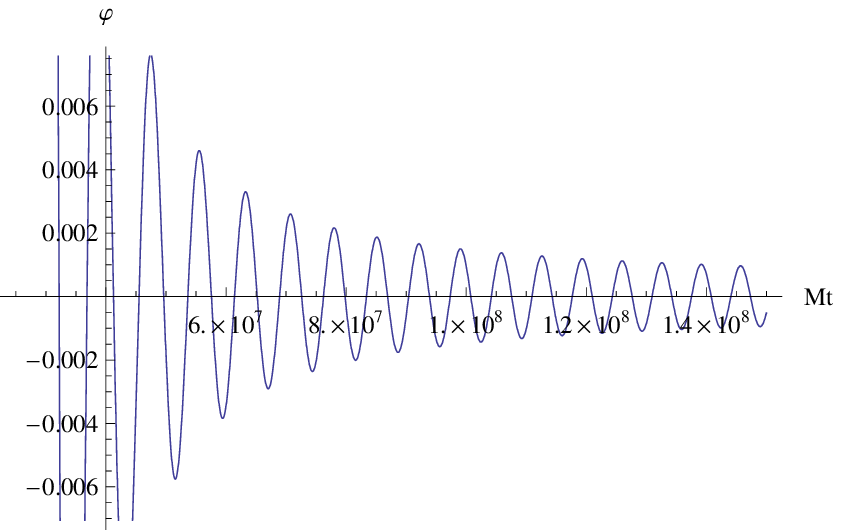}
\caption{Inflaton oscillations at the end of inflation. Values on vertical axes are expressed in Planck masses and time on horizontal axes is expressed in Planck times.}
\label{Pic2}
\end{minipage}
\end{figure}

\begin{figure}[h!]
\begin{minipage}[b]{0.47\linewidth}
\centering
\includegraphics[width=6 cm]{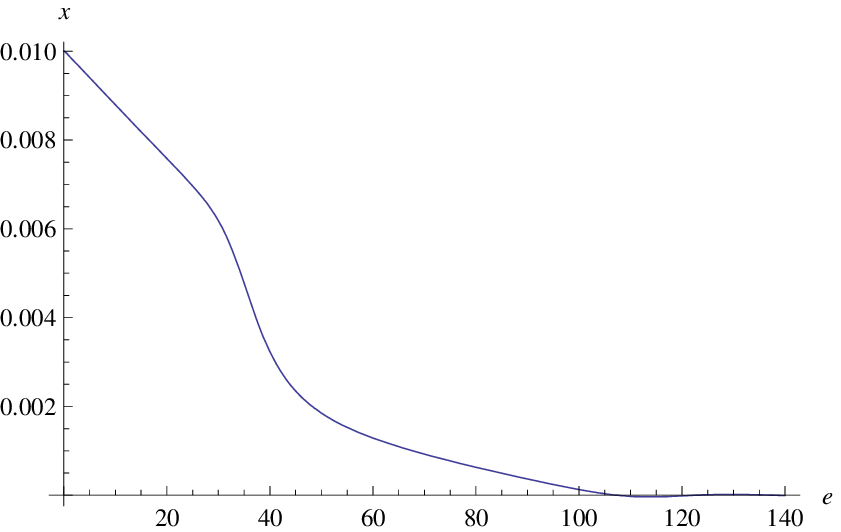}
\caption{Evolution of the field $x$ during inflation. Values on vertical axes are expressed in Planck masses and time on horizontal axes is expressed in the approximate number of e-folds of inflation.}
\label{Pic3}
\end{minipage}
\hspace{0.5cm}
\begin{minipage}[b]{0.47\linewidth}
\centering
\includegraphics[width=6 cm]{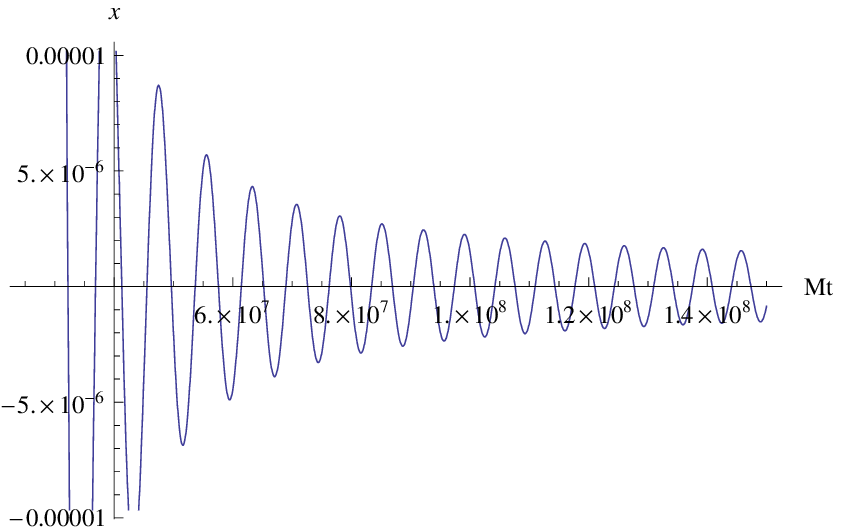}
\caption{Oscillations of the field $x$ at the end of inflation. Values on vertical axes are expressed in Planck masses and time on horizontal axes is expressed in Planck times.}
\label{Pic4}
\end{minipage}
\end{figure}

\begin{figure}[h!]
\centering
\includegraphics[width=8 cm]{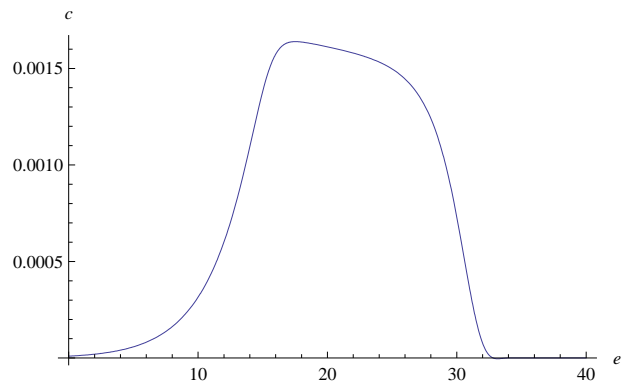}
\caption{Evolution of the field $c$ during during inflation. Values on vertical axes are expressed in Planck masses and time on horizontal axes is expressed in the approximate number of e-folds of inflation.}
\label{Pic5}
\end{figure}

\begin{figure}[h!]
\begin{minipage}[b]{0.47\linewidth}
\centering
\includegraphics[width=8 cm]{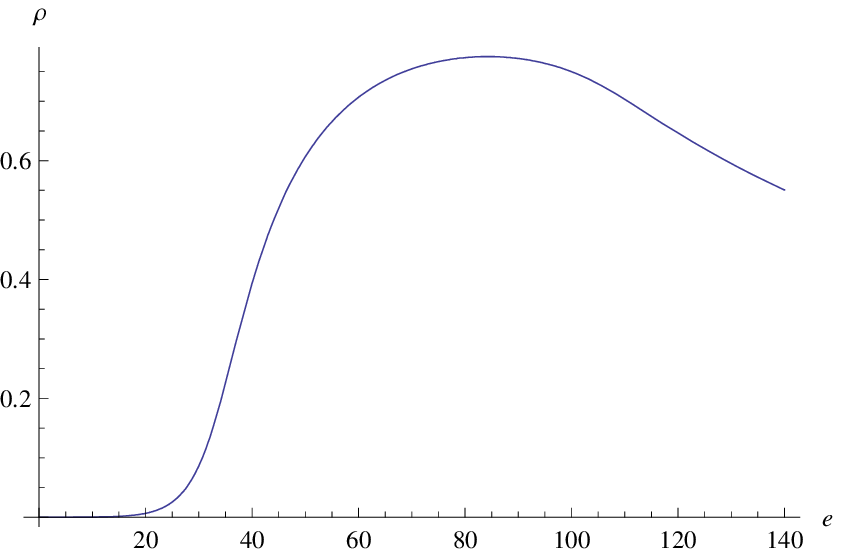}
\caption{Evolution of the field $\rho$ during inflation. Values on vertical axes are expressed in Planck masses and time on horizontal axes is expressed in the approximate number of e-folds of inflation.}
\label{p8}
\end{minipage}
\hspace{0.5cm}
\begin{minipage}[b]{0.47\linewidth}
\centering
\includegraphics[width=8 cm]{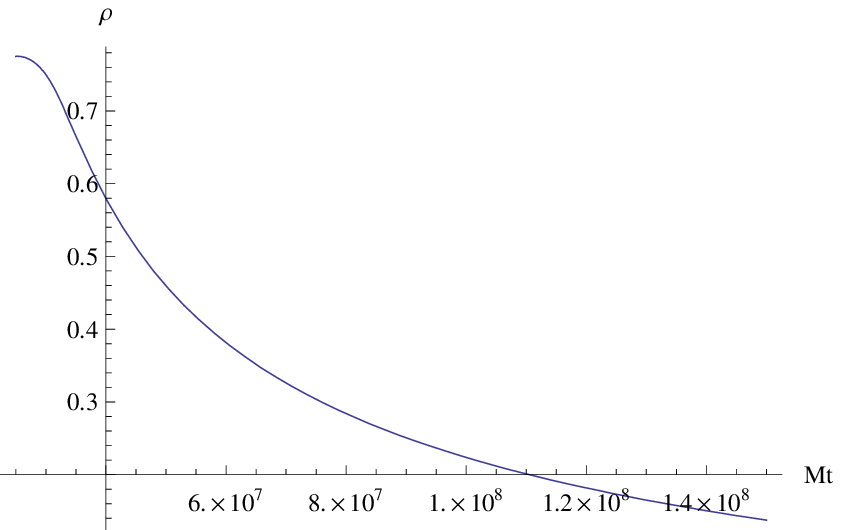}
\caption{Evolution of the field $\rho$ at the end of inflation. Values on vertical axes are expressed in Planck masses and time on horizontal axes is expressed in Planck times.}
\label{Pic6}
\end{minipage}
\end{figure}

\begin{figure}[h!]
\centering
\includegraphics[width=8 cm]{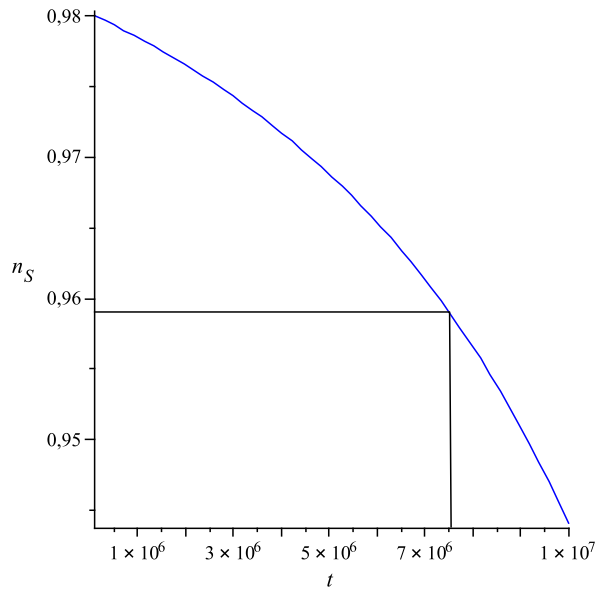}
\caption{Evolution of the spectral index $n_{S}$ between 100 and 40 e-folds before the end of inflation. Black lines marks the time of 50 e-folds before the end of inflation and corresponding value of the spectral index. Time on horizontal axes is expressed in Planck times.}
\label{Pic7}
\end{figure}

Preliminary numerical calculations show that in this case some of remaining 7 real VEVs (after setting $\eta=0$) can be also neglected. First of all the scalar potential does not depend on the phase $\sigma$ and any initial velocity of this phase will fall to zero due to the Hubble friction term. Therefore the flat direction $udd$ can be effectively described by only its absolute value $\rho$. The VEV of the field $c$, the absolute value of Higgs flat direction, after an initial increase, falls to zero quite rapidly. Thereafter the scalar potential ceases to depend on $\beta$ and $\kappa$.\newline
Henceforth we present more accurate calculations only for 4 real fields: $\varphi$, $x$, $c$ and $\rho$. All three phases have been set to zero. In this scenario the final 100 e-folds of inflation are studied.\newline
The inflaton VEV evolves according to standard chaotic inflation. It smoothly decreases during inflation and starts to oscillate after its end, as can be seen in figures (\ref{Pic1}) and (\ref{Pic2}). 
The field $x$ behaves almost identically to the inflaton field $\varphi$. The main difference is that the VEV of the field $x$ is much smaller than the VEV of the inflaton. This is illustrated in figures (\ref{Pic3}) and (\ref{Pic4}).
The VEV of the field $c$ initially rises due to the influence of the non-minimal K\"{a}hler coupling which shifts the minimum of the scalar potential for this direction away from zero. After the initial rise this VEV starts decreasing and drops to zero rapidly during the first 35 e-folds of inflation. This effect is shown in figure (\ref{Pic5}) and is caused by the relatively large value of $\lambda_{\chi}$ ($\lambda_{\chi}=1$) with respect to $\lambda_{\alpha}$, which favors the creation of large $udd$ direction VEV. Due to the Yukawa coupling between $udd$ and $H_{u}H_{d}$ directions a large $udd$ direction VEV induces an effective mass for the $H_{u}H_{d}$ direction. This effect shifts the minimum of the scalar potential for the $H_{u}H_{d}$ direction to zero leading to the rapid decrease of the $H_{u}H_{d}$ VEV during inflation.\newline
The VEV of the field $\rho$ rises during inflation to a value which is close to Planck mass due to the small value of $\lambda_{\alpha}$ (Fig. (\ref{p8})). At the end of inflation it starts to slowly decrease, as illustrated in Fig. (\ref{Pic6}). 
The evolution of the spectral index in the crucial period 60-50 e-folds before the end of inflation can be calculated in the slow-roll regime and is shown in figure (\ref{Pic7}).
The value of the spectral index 50 e-folds before the end of inflation is in agreement with the value derived from the WMAP5 data $n_{s}=0.960^{+0.014}_{-0.013}$ \cite{komatsu}.

\subsection{$\lambda_{\alpha}=1$ and $\lambda_{\chi}=1$}

\begin{figure}[h!]
\centering
\includegraphics{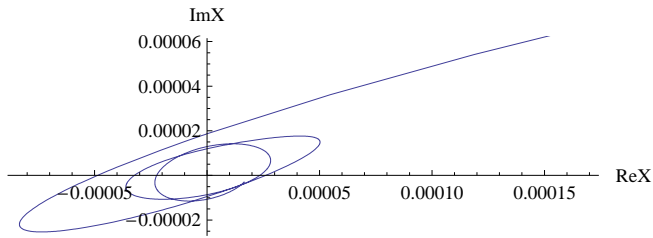}
\caption{Evolution of field $X$ on the complex plane during inflaton oscillations.}
\label{p37}
\end{figure}

\begin{figure}[h!]
\begin{minipage}[b]{0.47\linewidth}
\centering
\includegraphics[width=8 cm]{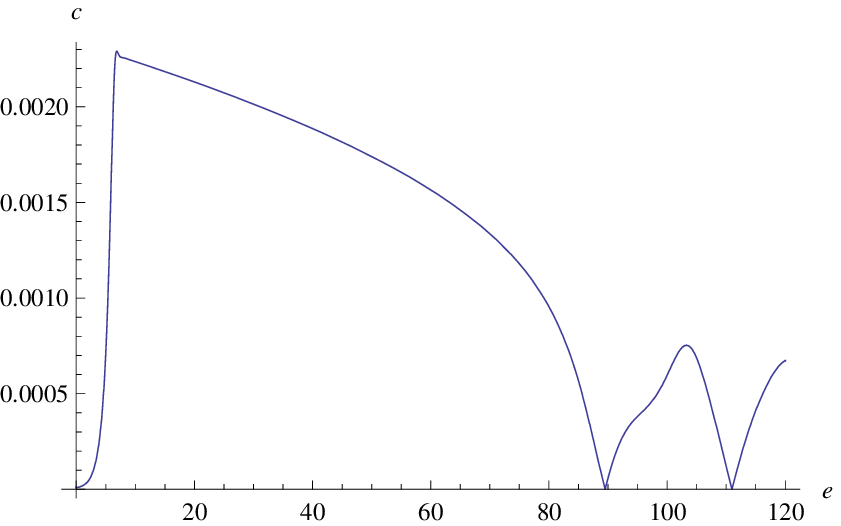}
\caption{Evolution of the absolute value of the Higgs field direction during inflation. Values on vertical axes are expressed in Planck masses and time on horizontal axes is expressed in the approximate number of e-folds of inflation.}
\label{p40}
\end{minipage}
\hspace{0.5cm}
\begin{minipage}[b]{0.47\linewidth}
\centering
\includegraphics[width=8 cm]{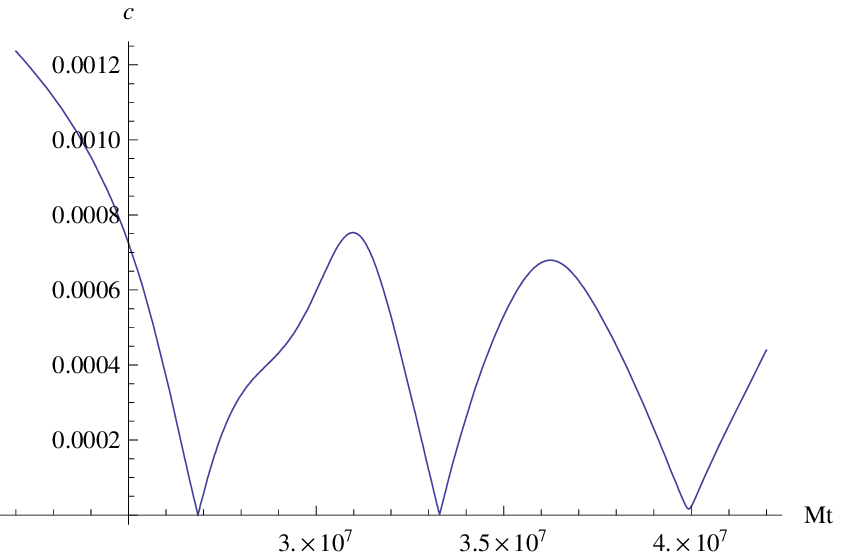}
\caption{Evolution of the absolute value of the Higgs field direction during inflaton oscillations. Values on vertical axes are expressed in Planck masses and time on horizontal axes is expressed in Planck times.}
\label{p41}
\end{minipage}
\end{figure}

\begin{figure}[h!]
\centering
\includegraphics{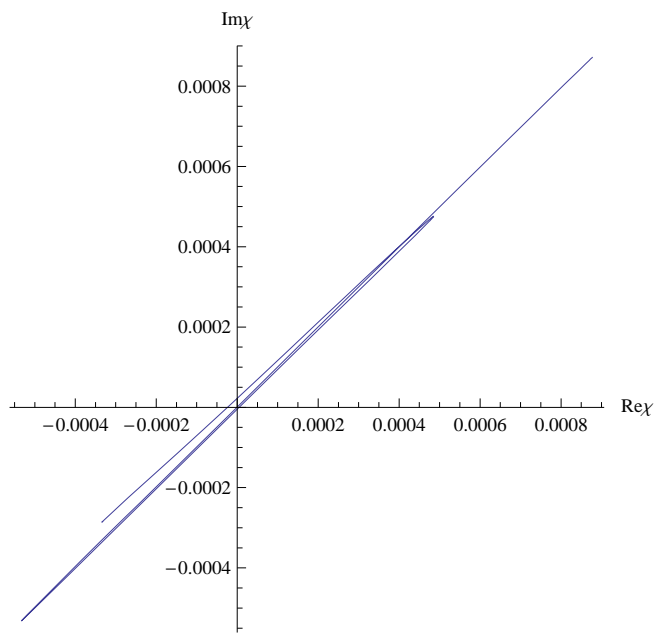}
\caption{Evolution of the Higgs field direction in the complex plane during inflaton oscillations}
\label{p42}
\end{figure}

\begin{figure}[h!]
\begin{minipage}[b]{0.47\linewidth}
\centering
\includegraphics[width=8 cm]{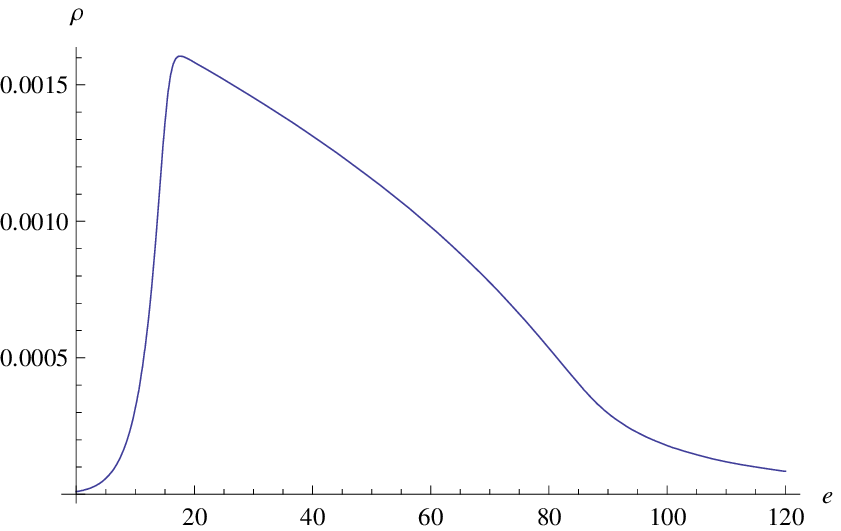}
\caption{Evolution of the absolute value of the $udd$ flat direction during inflation. Values on vertical axes are expressed in Planck masses and time on horizontal axes is expressed in the approximate number of e-folds of inflation.}
\label{p38}
\end{minipage}
\hspace{0.5cm}
\begin{minipage}[b]{0.47\linewidth}
\centering
\includegraphics[width=8 cm]{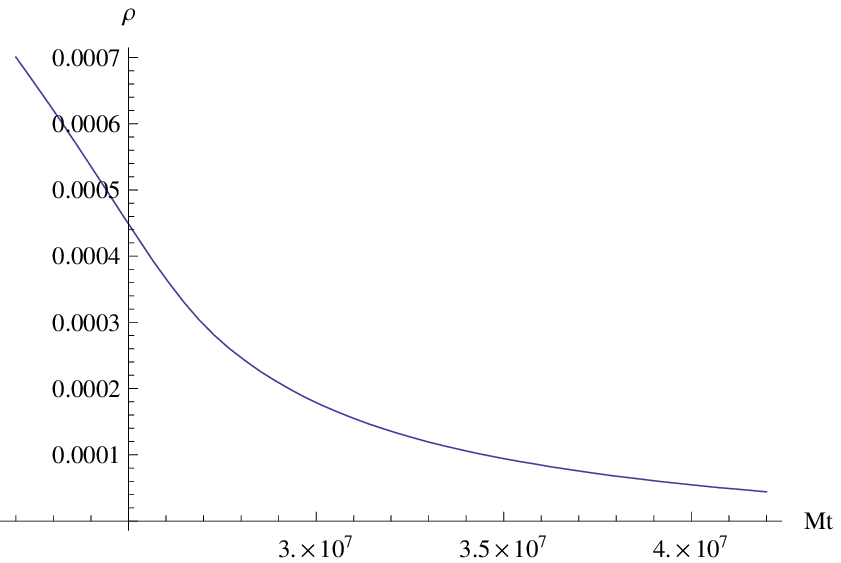}
\caption{Evolution of the absolute value of the $udd$ flat direction during inflaton oscillations. Values on vertical axes are expressed in Planck masses and time on horizontal axes is expressed in Planck times.}
\label{p39}
\end{minipage}
\end{figure}

The inflaton VEV evolves according to standard chaotic inflation as in the previous case. Its evolution includes the slow-roll period which naturally ends with inflaton oscillations.
Due to the dependence of the scalar potential on the phases of fields $X$ and $\chi$, both of these fields evolve non-trivially in the complex plane. The evolution of the absolute value of field $X$ is similar to the previous scenario and mimics the behavior of the inflaton. Figure (\ref{p37}) shows the evolution of field $X$ in the complex plane.
Since $\lambda_{\alpha}\sim\lambda_{\chi}$ now the absolute values of fields corresponding to both flat directions evolve similarly. Numerical calculations show that the absolute values of both fields grow during inflation, achieving a maximum of the order of $10^{-3}M_{Pl}$ and then start to decrease at the end of inflation due to the time-evolution of the minimum of their potentials. Because of the dynamics of the phase of field $\chi$ the effective mass of the field is slightly different to that of the field $\alpha$. As a result field $\chi$ begins oscillating earlier. Figures (\ref{p40}) and (\ref{p41}) show the evolution of the absolute value of $\chi$, while figure (\ref{p42}) shows the evolution of $\chi$ in the complex plane.
The evolution of the $udd$ flat direction is effectively one-dimensional because the scalar potential does not depend on the phase of this field and the Hubble friction quickly suppresses any initial phase dynamics. Figures (\ref{p38}) and (\ref{p39}) show the evolution of the absolute value of $\alpha$. 
Using the numerically found evolution the Hubble parameter and slow-roll parameters can be calculated. They fulfill the slow-roll conditions. One can then obtain the spectral index in the slow-roll approximation. The spectral index evaluated at the time of about 50-60 e-folds before the end of inflation is consistent with the value of the spectral index derived from the WMAP5 observational data \cite{komatsu}.

\section{Excitations around VEVs}

In this chapter we introduce excitations around all MSUGRA fields, which are related to the flat directions under consideration. By "related" we mean that either they have large VEVs, parametrized by flat directions, or they are parts of multiplets in which other fields have such VEVs. In the first case we parametrize those fields in the following way
\begin{equation}
F_{1}=\left(\frac{|VEV|}{\sqrt{n}}+\frac{\xi_{1}}{\sqrt{2}}\right)e^{i\left(arg\left(VEV\right)+\frac{\xi_{2}}{\sqrt{2}|VEV|}\right)}.
\label{f1}
\end{equation}
In the second case the parametrization has the form
\begin{equation}
F_{2}=\frac{1}{2}\left(\xi_{3}+i\xi_{4}\right)e^{i\cdot arg\left(VEV\right)}.
\label{f2}
\end{equation}
In formulas (\ref{f1}) and (\ref{f2}) $\xi_{1}$, $\xi_{2}$, $\xi_{3}$ and $\xi_{4}$ are real excitations, whereas $VEV$ denotes $\chi$ or $\alpha$. Moreover $n=2$ for Higgs doublets and $n=3$ for squark triplets. If the VEV of a flat direction drops to zero, the parametrization of excitations around the related MSUGRA fields is straightforward
\begin{equation}
F_{3}=\frac{1}{\sqrt{2}}\left(\delta_{1}+i\delta_{2}\right).
\label{f3}
\end{equation}
Excitations around fields $\Phi$ and $X$ are not considered. \newline We initially have 26-dimensional space of real excitations. Some are Goldstone bosons related to the gauge group generators which are broken by the VEVs of flat direction(s). Goldstone bosons can be eliminated via the Higgs mechanism. Then, expanding the Lagrangian density in the remaining excitations
\begin{equation}
L=L_{0}+L_{1}+L_{2}+\ldots
\label{Lexpansion}
\end{equation}
Here $L_{n}$ is the part of Lagrangian density which is of n-th order in the excitations. In particular $L_{0}$ is the classical and homogeneous limit of the Lagrangian density, used previously to obtain the evolution of VEVs. Terms $L_{n}$ for $n>2$ are neglected from now on. $L_{1}$ can be set to zero after using partial integration of the action and the classical equations of motion. The remaining part $L_{2}$ is used to describe particle production on the classical, homogeneous background including the dynamics of VEVs and the scale factor. The kinetic part of $L_{2}$ is quite complicated due to the non-minimal part in the K\"{a}hler Potential $K_{NM}$. However, as was mentioned in section 3, excitations are considered only at the end of inflation and later. In this period $K_{NM}$ can be neglected since the multiplicative factor $\frac{a}{M^{2}_{4}}XX^{*}$, included in equation (\ref{K_NM}), is very small. This approximation is adopted only for kinetic terms, keeping the full SUGRA scalar potential (we checked numerically that using minimal kinetic terms during inflaton oscillations does not change the classical evolution of the fields, while changing the potential does alter the evolution). Then, $L_{2}$ has the following general form
\begin{equation}
L_{2}=\frac{1}{2}\partial_{\mu}\Xi^{T}\partial^{\mu}\Xi-\frac{1}{2}\Xi^{T}M^{2}\Xi-\dot{\Xi}^{T}U\Xi.
\label{L_2}
\end{equation}
Here $\Xi$ is the vector which contains all excitations and $M^{2}$ is the mass matrix. There is also a matrix $U$, which mixes excitations with their derivatives. Elements of the matrices $M^{2}$ and $U$ can in general be functions of all 8 real VEVs and the matrix $U$ can also contain derivatives of those VEVs. There are two steps needed to transform $L_{2}$ into more convenient form:
\begin{enumerate}
\item Integrating the term $\dot{\Xi}^{T}U\Xi$ by parts in the action in order to replace the matrix $U$ with an antisymmetric matrix $\hat{U}$. This procedure gives new contributions to the mass matrix. Its new form will be denoted by $\hat{M}^{2}$
\begin{equation}
L_{2}=\frac{1}{2}\partial_{\mu}\Xi^{T}\partial^{\mu}\Xi-\frac{1}{2}\Xi^{T}\hat{M}^{2}\Xi-\dot{\Xi}^{T}\hat{U}\Xi.
\label{L_2a}
\end{equation}
\item Since $\hat{U}$ is antisymmetric, one can find an orthogonal matrix $A$ such that
\begin{equation}
\hat{U}=\dot{A}^{T}A.
\label{A}
\end{equation}
Defining $\Xi'\equiv A\Xi$, the $\hat{U}$ matrix can be eliminated
\begin{equation}
L_{2}=\frac{1}{2}\partial_{\mu}\Xi'^{T}\partial^{\mu}\Xi'-\frac{1}{2}\Xi'^{T}M'^{2}\Xi',
\label{L_2b}
\end{equation}
where $M'^{2}\equiv A\left(\hat{M}^{2}-\hat{U}^{2}\right)A^{T}$. 
\end{enumerate}
Similar transformations of $L_{2}$ were presented in \cite{basboll}. The matrix $M'^{2}$ can be diagonalized
\begin{equation}
M'^{2}=C M^{2}_{d}C^{T},
\label{M_diag}
\end{equation}
where the matrix $M^{2}_{d}$ is diagonal and the matrix $C$ is orthogonal. The situation is more general than in \cite{basboll}, because both these matrices are functions of VEVs, so both of them depend on time. We have time dependent eigenvectors, but also time dependent eigenvalues of matrix $M'^{2}$ and both these time dependences influence particle production.
\newline\newline To calculate particle production we quantize excitations in curved space-time according to \cite{Ford}. Quantum excitations are minimally coupled to gravity (they don't have any couplings to the Ricci scalar) and have effective squared masses which may differ significantly from the eigenvalues of $M'^{2}$, according to the following formula
\begin{equation}
m^{2}_{eff \xi}=m^{2}_{\xi}-2H^{2}-\dot{H}.
\label{effmass}
\end{equation}
Here $m^{2}_{\xi}$ is the ordinary squared mass of an excitation $\xi$ (an eigenvalue of $M'^{2}$), $m^{2}_{eff \xi}$ is the effective squared mass of this excitation and $H$ is the Hubble parameter.\newline Two sets of excitation modes are used so that the corresponding vacua minimize Hamiltonian in two particular moments of time. The set of $in$ modes minimize the Hamiltonian at $t_{0}$, the beginning of the considered particle creation period, and the set of $out$ modes minimize the Hamiltonian at $t_{1}$, the end of this period. After finding numerically the evolution of $in$ modes between $t_{0}$ and $t_{1}$ the Bogolyubov coefficients method \cite{bunch, birrell, Ford} is used to obtain the energy density of produced particles. It is worth noting that the above method allows us to describe particle production in the fully non-perturbative regime. We do not need space-time to be initially close to de Sitter and we do not use adiabatic modes. Non-adiabatic particle production is efficient when the adiabaticity condition (\ref{adiabat}) is broken, \cite{mazumdar}. Therefore in each of the cases, we choose a period of time for our calculations that corresponds to the broken adiabaticity condition.

\subsection{$\lambda_{\alpha}=10^{-7}$ and $\lambda_{\chi}=1$}

\begin{figure}[h!]
\centering
\includegraphics{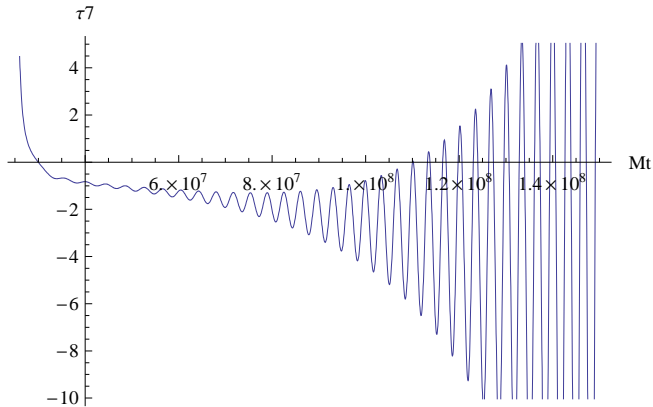}
\caption{Evolution of the adaibaticity parameter related to $m_{\xi_{2}}$}
\label{p23}
\end{figure}

\begin{figure}[h!]
\centering
\includegraphics[width=8 cm]{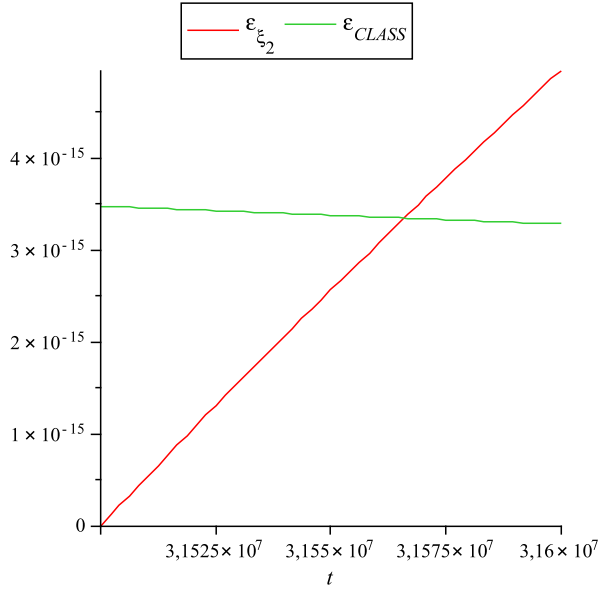}
\caption{Energy density of created particles $\epsilon_{\xi_{2}}$ in comparison to the total classical energy density of all VEVs $\epsilon_{CLASS}$. Time on horizontal axes is expressed in Planck times.}
\label{Pic12}
\end{figure}

In this case the only flat direction which still has a VEV at the end of inflation is the $udd$ direction. Therefore excitations which correspond to Higgs doublets are described by formula (\ref{f3}). Other excitations, which correspond to squark triplets, are described by formulae (\ref{f1}) and (\ref{f2}). The VEV of the field $\rho$ breaks the gauge symmetry
\begin{equation}
SU(3)_{C}\times U(1)_{Y} \rightarrow U(1)_{P}.
\label{SU}
\end{equation}
Here $U(1)_{P}$ is parametrized by a single generator $P$, which is defined in the following way
\begin{equation}
P=Y-\frac{1}{2}J_{3}-\frac{\sqrt{3}}{6}J_{8},
\label{PP}
\end{equation}
where $Y$ is the weak hypercharge, while $J_{3}$ and $J_{8}$ are two $SU(3)_{C}$ generators, which can be represented as two diagonal Gel-Mann matrices. The $SU(2)_{L}$ symmetry remains unbroken. There are 8 broken generators, related to 8 Goldstone bosons among 26 initial excitations. Hence there are 18 physical degrees of freedom in the excitation space. After eliminating Goldstone bosons, we use the unitary gauge to parametrize this 18-dimensional space. The excitations around fields in the Higgs doublets are
\begin{equation}
\begin{array}{clrr}%
H_{u\;1}=\frac{1}{\sqrt{2}}(\delta_{1}+i\delta_{2}), \\
H_{u\;2}=\frac{1}{\sqrt{2}}(\delta_{3}+i\delta_{4}), \\
H_{d\;1}=\frac{1}{\sqrt{2}}(\delta_{5}+i\delta_{6}), \\
H_{d\;2}=\frac{1}{\sqrt{2}}(\delta_{7}+i\delta_{8}).
\end{array}
\label{exHH}
\end{equation}
Excitations around fields in squark triplets in the unitary gauge (after eliminating Goldstone bosons) take the form\footnote{In section 3.1.$\sigma$ has been set to zero }
\begin{equation}
\begin{array}{clrr}%
u_{i\;1}=\left(\frac{\rho}{\sqrt{3}}+\frac{\xi_{7}}{\sqrt{2}}\right)e^{i\left(\sigma+\frac{\xi_{2}}{\sqrt{2}\rho}\right)}, \\
u_{i\;2}=\frac{1}{2}\left(\xi_{8}+i\xi_{9}\right)e^{i\sigma}, \\
u_{i\;3}=\frac{1}{2}\left(\xi_{10}+i\xi_{11}\right)e^{i\sigma}, \\
d_{j\;1}=\frac{1}{2}\left(\xi_{8}-i\xi_{9}\right)e^{i\sigma}, \\
d_{j\;2}=\left(\frac{\rho}{\sqrt{3}}+\frac{\xi_{12}}{\sqrt{2}}\right)e^{i\left(\sigma+\frac{\xi_{2}}{\sqrt{2}\rho}\right)}, \\
d_{j\;3}=\frac{1}{2}\left(\xi_{13}+i\xi_{14}\right)e^{i\sigma}, \\
d_{k\;1}=\frac{1}{2}\left(\xi_{10}-i\xi_{11}\right)e^{i\sigma}, \\
d_{k\;2}=\frac{1}{2}\left(\xi_{13}-i\xi_{14}\right)e^{i\sigma}, \\
d_{k\;3}=\left(\frac{\rho}{\sqrt{3}}+\frac{\xi_{15}}{\sqrt{2}}\right)e^{i\left(\sigma+\frac{\xi_{2}}{\sqrt{2}\rho}\right)}.
\end{array}
\label{exuud}
\end{equation}
We followed the procedure described earlier in this section and found that the $\hat{U}$ matrix vanishes in this case, because there is no phase dynamics, so $M'^{2}=\hat{M}^{2}$. The mass matrix of the excitations related to Higgs fields and $udd$ fields has a block diagonal form
\begin{equation}
M'^{2}=\left(
\begin{matrix}
M^{2}_{8\times8}\left[H_{u}H_{d}\right] &  0 \\
0 & M^{2}_{10\times10}\left[udd\right]
\end{matrix}
\right).
\label{mpr2}
\end{equation}
The $8\times8$-dimensional block of Higgs-related excitations has 8 eigenvalues, which (under the simplifying assumption that the Yukawa matrix is diagonal in flavor and $\lambda_{u}=\lambda_{d}=Y$) are degenerate. Four of them have the following approximate form
\begin{equation}
m^{2}_{1}\approx-\frac{m\varphi}{2}\left(2\sqrt{2}h+\left(a-1\right)m\varphi\right)
\nonumber
\end{equation}
\begin{equation}
+\left(\frac{Y^{2}}{3}+\sqrt{2}\left(a-1\right)hm\varphi+\left(1-2a+2a^{2}\right)\frac{m^{2}\varphi^{2}}{2}\right)\rho^{2}+O\left[x^{2},\;x^{2}\rho^{2},\;\rho^{3}\right],
\end{equation}
while the other four are given by the following approximate expression:
\begin{equation}
m^{2}_{2}\approx-\frac{m\varphi}{2}\left(-2\sqrt{2}h+\left(a-1\right)m\varphi\right)
\nonumber
\end{equation}
\begin{equation}
+\left(\frac{Y^{2}}{3}-\sqrt{2}\left(a-1\right)hm\varphi+\left(1-2a+2a^{2}\right)\frac{m^{2}\varphi^{2}}{2}\right)\rho^{2}+O\left[x^{2},\;x^{2}\rho^{2},\;\rho^{3}\right],
\end{equation}
where we have neglected the soft masses and the Higgs mass parameter $\mu$. In the equations above it can be seen that there are, as expected, two main contributions to the mass eigenvalues. One comes from the standard 3- and 4-linear interaction terms with the inflaton. The other is related to the flat direction and the dominant contribution comes from the Yukawa interaction between Higgs fields and the $udd$ flat direction fields. The first term in each parenthesis, scaled by $h$ or $Y$, is related to the assumed couplings in the superpotential and would survive even in the absence of gravitational effects. All the additional terms come from supergravity and the parameter $a$ scales the influence of the non-minimal coupling in the K\"{a}hler potential. Due to the Yukawa coupling the influence of the $udd$ flat direction VEV on the mass eigenvalues related to Higgs fields dominates. Therefore these eigenvalues are large and evolve slowly in time, so preheating from the inflaton into particles related to those eigenvalues is initially blocked.\newline
The $10\times10$-dimensional block of the mass matrix, which is related to $udd$ excitations, is also block-diagonal. It has one 3-dimensional block and seven 1-dimensional blocks. 
\begin{equation}
M^{2}\left[udd\right]=\left(
\begin{matrix}
M^{2}_{1\times1}\left[phase\right] & 0 & 0 & 0 & 0 & 0 & 0 & 0 \\
0 & M^{2}_{3\times3}\left[flat\right] & 0 & 0 & 0 & 0 & 0 & 0 \\
0 & 0 & M^{2}_{1\times1}\left[1\right] & 0 & 0 & 0 & 0 & 0 \\
0 & 0 & 0 & M^{2}_{1\times1}\left[2\right] & 0 & 0 & 0 & 0 \\
0 & 0 & 0 & 0 & M^{2}_{1\times1}\left[3\right] & 0 & 0 & 0 \\
0 & 0 & 0 & 0 & 0 & M^{2}_{1\times1}\left[4\right] & 0 & 0 \\
0 & 0 & 0 & 0 & 0 & 0 & M^{2}_{1\times1}\left[5\right] & 0 \\
0 & 0 & 0 & 0 & 0 & 0 & 0 & M^{2}_{1\times1}\left[6\right]
\end{matrix}
\right).
\end{equation}
Eigenvalues $M^{2}\left[1\right]$ - $M^{2}\left[6\right]$ correspond to combinations of excitations around fields with $VEV=0$ belonging to $udd$ flat direction multiplets. They are all heavy because they are related to Higgs particles coming from broken non-diagonal generators of SU(3). For example (under the simplifying assumption that all the gauge couplings are equal, $g_{i}=g$)
\begin{equation}
M^{2}\left[1\right]\approx-\frac{m^{2}\varphi^{2}}{2}\left(a-1\right)+\left(\frac{g^{2}}{3}+\left(1-2a+2a^{2}\right)\frac{m^{2}\varphi^{2}}{2}+O\left[x^{2}\right]\right)\rho^{2}+O\left[x^{2},\;\rho^{3}\right].
\end{equation}
It can be observed that the contribution to this eigenvalue coming from the inflaton VEV is a supergravity effect. This is true for all the eigenvalues of $M^{2}\left[udd\right]$ - in the global SUSY case $udd$ fields are not coupled to the inflaton. The dominant contribution to $M^{2}\left[1\right]$ comes from D-terms, is scaled by the gauge coupling $g$ and is proportional to the VEV$^{2}$ of the flat direction. Therefore all the eigenvalues $M^{2}\left[1\right]$ - $M^{2}\left[6\right]$ related to broken non-diagonal generators are heavy and evolve slowly in time. At the beginning of inflation there is no non-perturbative particle production of particles corresponding to those eigenvalues. The block $M^{2}\left[flat\right]$ corresponds to excitations $\xi_{7},\ \xi_{12}$ and $\xi_{15}$ around the absolute value of the VEV of flat direction $udd$. It can be diagonalized easily, giving two heavy eigenvalues and one light eigenvalue. Both heavy eigenvalues are dominated by terms $\sim{}g^{2}\rho^{2}$ and are Higgs particles corresponding to two broken diagonal generators of SU(3). As in previous cases preheating into those particles is blocked. The light eigenvalue $m^{2}_{3}$ corresponds to the combination $\left(\xi_{7}+\xi_{12}+\xi_{15}\right)/\sqrt{3}$ of excitations and is given approximately by
\begin{equation}
m^{2}_{3}\approx-\frac{m^{2}\varphi^{2}}{2}\left(a-1\right)+3\left(1-2a+2a^{2}\right)\frac{m^{2}\varphi^{2}}{2}\;\rho^{2}+O\left[x^{2},\;x^{2}\rho^{2},\;\rho^{3}\right].
\label{m3}
\end{equation}
Clearly this eigenvalue is dominated by supergravity effects (all terms written explicitly in eq. (\ref{m3}) are induced by supergravity) - the only contribution which would remain in global SUSY comes from the non-renormalizable term and is of the order $\sim\rho^{4}$. In global SUSY without non-renormalizable terms this eigenvalue would be equal to zero. This is easy to understand since in global SUSY without non-renormalizable terms the scalar potential does not depend on the flat direction. Then there exist two massless eigenvalues corresponding to excitations around the complex field $\alpha$ parameterizing the flat direction. When the scalar potential depends on the flat direction (due to non-renormalizable terms or supergravity effects), those two states gain mass. The eigenvalue $m^{2}_{3}$ corresponds to one of such states. The mass eigenvalues corresponding to excitations around $\alpha$ remain naturally light, because any contributions to them originate from non-renormalizable or supergravity terms and are diminished by the Planck scale. Moreover, as can be seen from (\ref{m3}), the influence of the inflaton VEV on these eigenvalues can easily become dominant. Numerical calculations show that the time evolution of this eigenvalue allows for both efficient preheating from the inflaton and efficient non-perturbative particle production from the flat direction into excitations around the flat direction. Without supergravity corrections or non-renormalizable terms this effect would not be possible - the discussed mass matrix eigenvalue would be equal to zero. Adding a non-renormalizable term for the flat direction in global SUSY would make the eigenvalue non-zero and equal $15\lambda_{\rho}^{2}\rho^{4}$. Such an eigenvalue is light and dependent only on the flat direction VEV. This would lead to non-perturbative particle production from the flat direction due to the time evolution of this eigenvalue. Adding supergravity effects couples the flat direction to the inflaton, allowing for non-perturbative particle production from the inflaton as well.\newline
The second light eigenvalue of the mass matrix, which would be equal to zero without supergravity corrections or non-renormalizable terms, is $M^{2}\left[phase\right]$ and is related to the excitation $\xi_{2}$ around the $\sigma$ phase\footnote{Initially, there are 3 excitations around $\sigma$, but two of them are Goldstone bosons.}. 
The squared mass $M^{2}\left[phase\right]=m_{\xi_{2}}^{2}$ (an element and eigenvalue of matrix $M'^{2}$) has the following form
\begin{equation}
m^{2}_{\xi_{2}}=-\frac{\ddot{\rho}+3H\dot{\rho}}{\rho}=\frac{V,_{\rho}}{2\rho}
\label{exsigma}
\end{equation}
\begin{equation}
\approx\left(1-a\right)\frac{m^{2}\varphi^{2}}{2}+\left(1-2a+2a^{2}\right)\frac{m^{2}\varphi^{2}}{2}\;\rho^{2}+O\left[x^{2},\;x^{2}\rho^{2},\;\rho^{3}\right],
\end{equation}
where $H$ is the Hubble parameter. Due to the simple form of this mass eigenvalue, which comes only from the contribution of the kinetic terms to the mass matrix, one can write this eigenvalue explicitly. The mass eigenvalue $m^{2}_{\xi_{2}}$ is the smallest of the eigenvalues of the matrix $M'^{2}$. It can be observed that the eigenvalue is dominated by supergravity terms. The non-renormalizable term gives a contribution $3\lambda_{\rho}^2\rho^{4}$. In global SUSY without non-renormalizable terms this eigenvalue would be identically equal to zero. Evolution of the adiabaticity parameter (\ref{adiabat}) corresponding to $m_{\xi_{2}}$ at the end of inflation is shown in Fig. (\ref{p23}). 
It can be seen that the behaviour of the adiabaticity parameter becomes quickly dominated by the influence of the inflaton oscillating VEV. This leads to the breaking of the adiabaticity condition.
\newline To calculate production of particles one has to derive an effective mass squared according to equation (\ref{effmass}). 
It is different from $m^{2}_{\xi_{2}}$ because for such a small eigenvalue the influence of the evolving background is relevant. The adiabaticity condition is still broken for $m^{2}_{eff \xi_{2}}$. The effective mass squared $m^{2}_{eff \xi_{2}}$ is not always positive, but in the Bogolyubov coefficients method \cite{bunch, birrell, Ford} that we use for calculating particle production the effective squared mass has to be positive only at times $t_{0}$ and $t_{1}$. For the previously discussed light mass matrix eigenvalue (\ref{m3}) the adiabaticity parameter is smaller than for $m^{2}_{\xi_{2}}$ (though it rapidly becomes of order 1 during inflaton oscillations) and for all other eigenvalues of $M'^{2}$ (\ref{mpr2}) the adiabaticity parameter is smaller than $10^{-6}$. This result means that the excitation $\xi_{2}$ is the main channel for preheating. This channel does not appear in global supersymmetry without non-renormalizable terms \cite{olive, basboll}, because in such a framework the eigenvalue related to the excitation around the phase of the flat direction is identically massless. The calculated numerically energy density of produced particles - $\epsilon_{\xi_{2}}$ grows very rapidly and begins to be comparable to the total energy density of all classical VEVs - $\epsilon_{CLASS}$ as can be seen in figure (\ref{Pic12}).
The back-reaction of the produced particles on the classical VEVs evolution is not taken into account, so the result shown in figure (\ref{Pic12}) should be treated as an approximation only. However, from the approximate calculations it can be concluded that both non-perturbative particle production from the flat direction and preheating from the inflaton into particles corresponding to excitations $\xi_{2}$ and $\left(\xi_{7}+\xi_{12}+\xi_{15}\right)/\sqrt{3}$ is efficient enough in order to melt the flat direction VEV, which unblocks all other channels of preheating from the inflaton. Moreover, to obtain this result one needs only the mechanism of particle production due to changing mass matrix eigenvalues related to the flat direction. The mechanism of particle production due to changing mass matrix eigenvectors \cite{olive, basboll} is not necessary.\newline
Even though the phase dynamics is negligible in this scenario it is worth making one comment. In the diagonalizing base of excitations matrix $A$ appearing in eq. (\ref{A}), has a non-trivial form only for excitations $\xi_{2}$ and $\left(\xi_{7}+\xi_{12}+\xi_{15}\right)/\sqrt{3}$
\begin{equation}
A=\left(
\begin{matrix}
cos\sigma &  sin\sigma \\
-sin\sigma &  cos\sigma
\end{matrix}
\right).
\end{equation}
With $\sigma=const=0$ $A=1$ and so $M'^{2}=M^{2}$, the mass matrix eigenvectors are constant in time and there is no preheating of the type proposed in \cite{olive}. However if the phase dynamic were present, the $A$ matrix would mix in time two eigenstates with non-zero eigenvalues, which would lead to preheating from time-varying eigenvectors even from a single flat direction, in contradiction to \cite{olive} and \cite{basboll}. This happens due to the inclusion of supergravity corrections or non-renormalizable terms. Both these effects make the scalar potential dependent on the flat direction, which gives masses to eigenstates $\xi_{2}$ and $\left(\xi_{7}+\xi_{12}+\xi_{15}\right)/\sqrt{3}$. Without these effects, as in \cite{olive} and \cite{basboll}, eigenstates $\xi_{2}$ and $\left(\xi_{7}+\xi_{12}+\xi_{15}\right)/\sqrt{3}$ are massless and their mixing does not lead to particle production.

\subsection{$\lambda_{\alpha}=1$ and $\lambda_{\chi}=1$}

\begin{figure}[h!]
\centering
\includegraphics{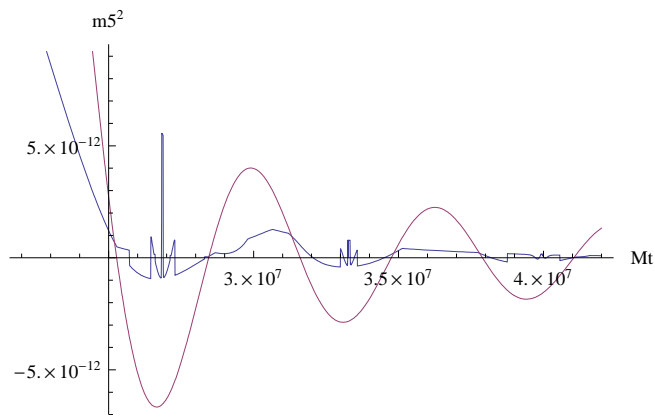}
\caption{Evolution of the light eigenvalue in comparison with the character of the evolution of the inflaton (red line)}
\label{p58}
\end{figure}

In this scenario both $udd$ and Higgs directions have non-zero VEVs during inflaton oscillations, breaking $SU(3)\times{}SU(2)\times{}U(1)\rightarrow{}U(1)$. After eliminating Goldstone bosons in the unitary gauge we are left with the following parametrization of excitations
\begin{equation}
\begin{array}{clrr}%
H_{u\;1}=\frac{1}{2}\left(\xi_{3}+i\xi_{4}\right)e^{i\kappa}, \\
H_{u\;2}=\left(\frac{c}{\sqrt{2}}+\frac{\xi_{5}}{\sqrt{2}}\right)e^{i\left(\kappa+\frac{\xi_{1}}{\sqrt{2}c}\right)}, \\
H_{d\;1}=\left(\frac{c}{\sqrt{2}}+\frac{\xi_{6}}{\sqrt{2}}\right)e^{i\left(\kappa+\frac{\xi_{1}}{\sqrt{2}c}\right)}, \\
H_{d\;2}=\frac{1}{2}\left(\xi_{3}-i\xi_{4}\right)e^{i\kappa}, \\
u_{i\;1}=\left(\frac{\rho}{\sqrt{3}}+\frac{\xi_{7}}{\sqrt{2}}\right)e^{i\left(\sigma+\frac{\xi_{2}}{\sqrt{2}\rho}\right)}, \\
u_{i\;2}=\frac{1}{2}\left(\xi_{8}+i\xi_{9}\right)e^{i\sigma}, \\
u_{i\;3}=\frac{1}{2}\left(\xi_{10}+i\xi_{11}\right)e^{i\sigma}, \\
d_{j\;1}=\frac{1}{2}\left(\xi_{8}-i\xi_{9}\right)e^{i\sigma}, \\
d_{j\;2}=\left(\frac{\rho}{\sqrt{3}}+\frac{\xi_{12}}{\sqrt{2}}\right)e^{i\left(\sigma+\frac{\xi_{2}}{\sqrt{2}\rho}\right)}, \\
d_{j\;3}=\frac{1}{2}\left(\xi_{13}+i\xi_{14}\right)e^{i\sigma}, \\
d_{k\;1}=\frac{1}{2}\left(\xi_{10}-i\xi_{11}\right)e^{i\sigma}, \\
d_{k\;2}=\frac{1}{2}\left(\xi_{13}-i\xi_{14}\right)e^{i\sigma}, \\
d_{k\;3}=\left(\frac{\rho}{\sqrt{3}}+\frac{\xi_{15}}{\sqrt{2}}\right)e^{i\left(\sigma+\frac{\xi_{2}}{\sqrt{2}\rho}\right)}.
\end{array}
\end{equation}
Due to a more complicated set of excitations in this scenario the eigenvectors and eigenvalues of the mass matrix have been found numerically. The mass matrix is block-diagonal - one complicated block is related to excitations around non-zero VEVs and a separate, diagonal block is related to excitations $\xi_{3},\ \xi_{4},\ \xi_{8},\ \xi_{9},\ \xi_{10},\ \xi_{11},\ \xi_{13}$ and $\xi_{14}$ around VEVs equal to zero.
\begin{equation}
M'^{2}=\left(
\begin{matrix}
M^{2}_{8\times8}\left[VEV\neq0\right] & 0 & 0 & 0 & 0 & 0 & 0 & 0 \\
0 & m^{2}_{I} & 0 & 0 & 0 & 0 & 0 & 0 \\
0 & 0 & m^{2}_{II} & 0 & 0 & 0 & 0 & 0 \\
0 & 0 & 0 & m^{2}_{III} & 0 & 0 & 0 & 0 \\
0 & 0 & 0 & 0 & m^{2}_{IV} & 0 & 0 & 0 \\
0 & 0 & 0 & 0 & 0 & m^{2}_{V} & 0 & 0 \\
0 & 0 & 0 & 0 & 0 & 0 & m^{2}_{VI} & 0 \\
0 & 0 & 0 & 0 & 0 & 0 & 0 & m^{2}_{VII}
\end{matrix}
\right)
\end{equation}
Mass matrix eigenvalues corresponding to $\xi_{3},\ \xi_{4},\ \xi_{8},\ \xi_{9},\ \xi_{10},\ \xi_{11},\ \xi_{13}$ and $\xi_{14}$ are heavy as they correspond to Higgs particles related to breaking of non-diagonal generators. Their time evolution is strongly dominated by the $udd$ flat direction or the Higgs direction VEVs - these eigenvalues are large and evolve slowly in time, which effectively blocks preheating of the inflaton into particles corresponding to these eigenvalues. The time evolution of some of the eigenvalues from the block corresponding to excitations around non-zero VEVs is also determined by large $udd$ and $H_{u}H_{d}$ VEVs. The evolution of these eigenvalues does not allow either non-perturbative particle production from the flat direction or preheating from the inflaton (these eigenvalues correspond mainly to excitations related to Higgs particles of the diagonal generators breaking). This type of behavior was predicted by ref. \cite{mazumdar}. Due to non-trivial phase dynamics the mass matrix eigenvectors also evolve in time allowing for non-perturbative particle production from flat directions as predicted in ref. \cite{olive}, which can lead to fast decay of flat direction VEVs. There is however another, more efficient channel of non-perturbative particle production due to the existence of light, non-adiabatically changing eigenvalues of the mass matrix. These light eigenvalues appear due to the same mechanism as described in the previous case - they correspond to a combination of naturally light excitations around VEVs of complex fields $\alpha$ and $\chi$ parameterizing the (quasi) flat directions. In this scenario however, due to the non-zero VEV of the $H_{u}H_{d}$ direction, preheating from the inflaton is allowed from the beginning of inflaton oscillations into excitations around both directions. As an example figure (\ref{p58}) shows the time evolution of such a light eigenvalue corresponding mainly to excitations around $H_{u}H_{d}$ direction. 
A comparison with the time evolution of the inflaton VEV clearly shows that the time evolution of this eigenvalue is dominated by the behavior of the inflaton. Moreover, because this eigenvalue is very small the impact of inflaton oscillations makes this eigenvalue periodically negative leading to a very effective tachyonic \cite{dufaux} preheating from the inflaton.

\section{Conclusions}

Achieving large flat direction VEVs through classical evolution during inflation is natural in a supergravity framework with non-minimal K\"{a}hler potential. Such large VEVs can block preheating from the inflaton into certain channels. However supergravity effects and non-renormalizable terms, which create a potential for the flat direction, are a source of light, rapidly changing eigenvalues of the mass matrix. They allow the non-perturbative production of particles from the flat direction and preheating from the inflaton. Non-zero VEVs of Higgs fields can also lead to the existence of light, rapidly evolving eigenvalues, allowing for preheating from the inflaton into Higgs particles from the beginning of inflaton oscillations. Non-perturbative particle production due to the time evolution of the mass matrix eigenstates is not necessary to reduce flat direction VEV and unblock preheating. Thus non-perturbative particle production from the inflaton is likely to remain the source of preheating even in the initial presence of large flat direction VEVs.

\section{Acknowledgements}

The authors would like to thank very much prof. Stefan Pokorski for all his help - looking over and inspiring their scientific work, correcting their mistakes, encouraging progress, discussing problems and always asking the most important questions.\newline
They would also like to thank prof. Keith Olive for an inspiring discussion and both prof. Keith Olive and prof. Marco Peloso for their kind interest and help with this work.\newline
Special thank you to Paul Hunt for his patient help with correcting the first version of this paper.


\begin{thebibliography}{23}
\bibitem{guth} A.H. Guth, \textit{The Inflationary Universe: A Possible Solution to the Horizon and Flatness Problems}, Phys.Rev.D23:347-356, 1981
\bibitem{linde1} A.D. Linde \textit{A New Inflationary Universe Scenario: A Possible Solution of the Horizon, Flatness, Homogeneity, Isotropy and Primordial Monopole Problems}, Phys.Lett.B108:389-393, 1982
\bibitem{albrecht} A. Albrecht, P.J. Steinhardt, \textit{Cosmology for Grand Unified Theories with Radiatively Induced Symmetry Breaking}, Phys.Rev.Lett.48:1220-1223, 1982
\bibitem{guth1} A.H. Guth, E.J. Weinberg, \textit{Could the Universe Have Recovered from a Slow First Order Phase Transition?}, Nucl.Phys.B212:321, 1983
\bibitem{mukhanov1} V.F. Mukhanov, \textit{Gravitational Instability of the Universe Filled with a Scalar Field}, JETP Lett. 41, 493 (1985) [Pisma Zh. Eksp. Teor. Fiz. 41, 402 (1985)]; M. Sasaki, \textit{Large Scale Quantum Fluctuations in the Inflationary Universe},
Prog. Theor. Phys. 76, 1036 (1986)
\bibitem{mukhanov2} V.F. Mukhanov, H.A. Feldman, R.H. Brandenberger, \textit{Theory of cosmological perturbations. Part 1. Classical perturbations. Part 2. Quantum theory of perturbations. Part 3. Extensions.}, Phys.Rept.215:203-333, 1992
\bibitem{colb} Edward W. Kolb, Michael S. Turner, \textit{The Early Universe}, Westview Press, 1990
\bibitem{mukhanov} Viatcheslav Mukhanov, \textit{Physical Foundations of Cosmology}, Cambridge University Press, 2005
\bibitem{albrecht1} A. Albrecht, P.J. Steinhardt, M.S. Turner, F. Wilczek, \textit{Reheating an Inflationary Universe}, Phys.Rev.Lett.48:1437, 1982
\bibitem{abbott} L.F. Abbott, E. Farhi, M.B. Wise, \textit{Particle Production in the New Inflationary Cosmology}, Phys.Lett.B117:29, 1982
\bibitem{dolgov} A.D. Dolgov, A.D. Linde, \textit{Baryon Asymmetry in Inflationary Universe}, Phys.Lett.B116:329, 1982
\bibitem{fayet} P. Fayet, S. Ferrara, \textit{Supersymmetry}, Phys.Rept.32:249-334, 1977
\bibitem{nilles} H.P. Nilles, \textit{Supersymmetry, Supergravity and Particle Physics}, Phys. Rep. 110 (1984), 1-162
\bibitem{haber} H.E. Haber, G.L. Kane \textit{The Search for Supersymmetry: Probing Physics Beyond the Standard Model}, Phys.Rept.117:75-263, 1985
\bibitem{bailin} D. Bailin, A. Love, \textit{Supersymmetric Gauge Theory and String Theory}, Institute of Physics Publishing 1994
\bibitem{ramond} P. Ramond, \textit{Journeys Beyond the Standard Model}, Westview Press 2004
\bibitem{binetruy} P. Binetruy \textit{Supersymmetry}, Oxford University Press 2006
\bibitem{gherghetta} T. Gherghetta, C. Kolda, S.P. Martin, \textit{Flat directions in the scalar potential of the supersymmetric standard model}, Nucl.Phys.B468:37-58, 1996, hep-ph/9510370
\bibitem{dine} M. Dine, L. Randall, S. Thomas, \textit{Baryogenesis from Flat Directions of the Supersymmetric Standard Model}, Nucl.Phys.B458:291-326, 1996, hep-ph/9507453
\bibitem{kasuyaa} S. Kasuyaa, M. Kawasaki \textit{Towards the robustness of the Affleck-Dine baryogenesis}, Phys.Rev.D74:063507, 2006, hep-ph/0606123
\bibitem{mazumdar} R. Allahverdi, A. Mazumdar, \textit{Reheating in supersymmetric high scale inflation}, Phys.Rev.D76:103526, 2007, hep-ph/0603244
\bibitem{olive} K.A. Olive, M. Peloso \textit{The Fate of SUSY Flat Directions and their Role in Reheating}, Phys.Rev.D74:103514, 2006, hep-ph/0608096
\bibitem{allahverdi} R. Allahverdi, A. Mazumdar \textit{Longevity of Supersymmetric Flat Directions}, hep-ph/0608296
\bibitem{basboll1} A. Basb\o{}ll, D. Maybury, F. Riva, S.M. West \textit{Non-Perturbative Flat Direction Decay}, hep-ph/0703015
\bibitem{basboll} A. Basb\o{}ll, \textit{SUSY Flat Direction Decay - the prospect of particle production and preheating}, arXiv:0801.0745
\bibitem{olive1} A. Emir G\"{u}mr\"{u}k\c{c}\"{u}o\u{g}lu, K.A. Olive, M. Peloso, M. Sexton \textit{The nonperturbative decay of SUSY flat directions}, Phys.Rev.D78:063512, 2008, arXiv:0805.0273
\bibitem{komatsu} E. Komatsu, J. Dunkley, M. R. Nolta, C. L. Bennett, B. Gold, G. Hinshaw, N. Jarosik, D. Larson, M. Limon, L. Page, D. N. Spergel, M. Halpern, R. S. Hill, A. Kogut, S. S. Meyer, G. S. Tucker, J. L. Weiland, E. Wollack, E. L. Wright, \textit{Five-year Wilkinson Microwave Anisotropy Probe (WMAP) Observations: Cosmological Interpretation}, arXiv:0803.0547
\bibitem{kawasaki} M. Kawasaki, M. Yamaguchi, T. Yanagida, \textit{Natural Chaotic Inflation in Supergravity}, Phys.Rev.Lett.85:3572-3575, 2000, hep-ph/0004243
\bibitem{bagger} J. Bagger, E. Poppitz, L. Randall, \textit{Destabilizing divergences in supergravity theories at two loops}, Nucl.Phys.B455:59-82, 1995, hep-ph/950524
\bibitem{gaillard} M.K. Gaillard, V. Jain, \textit{Supergravity coupled to chiral matter at one loop}, Phys.Rev.D49:1951-1965, 1994, hep-ph/9308090
\bibitem{kawasaki1} M. Kawasaki, M. Yamaguchi, T. Yanagida, \textit{Natural Chaotic Inflation in Supergravity and Leptogenesis}, Phys.Rev.D63:103514, 2001, hep-ph/0011104
\bibitem{bunch} T. S. Bunch and P. C. W. Davies, Proc. R. Soc. London, Ser. A A360, 117 (1978)
\bibitem{birrell} N.D. Birrell, P.C.W. Davies, \textit{Quantum fields in curved space}, Cambridge University Press, 1982
\bibitem{Ford} L. H. Ford \textit{Quantum Field Theory In Curved Spacetime}, TUTP-97-9 To be published in the proceedings of 9th Jorge Andre Swieca Summer School: Particles and Fields, Sao Paulo, Brazil 1997, gr-qc/9707062 
\bibitem{dufaux} J. F. Dufaux, G. N. Felder, L. Kofman, M. Peloso, D. Podolsky, \textit{Preheating with Trilinear Interactions: Tachyonic Resonance}, JCAP 0607:006, 2006, hep-ph/0602144
\end{thebibliography}
\end {document}